\pdfoutput=1
\documentclass[aps,prl,twocolumn,showpacs,psfig,superscriptaddress
]{revtex4-1}
\usepackage{graphicx}% Include figure files
\usepackage{dcolumn}% Align table columns on decimal point
\usepackage{bm}% bold math
\usepackage{amsmath}
\usepackage{lmodern}

\usepackage{color}
\usepackage[dvipsnames]{xcolor}
\usepackage{amssymb}
\usepackage{braket}
\usepackage{mathtools}
\usepackage[colorlinks=true,linkcolor=blue,citecolor=blue,urlcolor=blue]{hyperref}
\usepackage{breakurl}
\usepackage{ulem}
\usepackage{graphicx}
\usepackage{lmodern}
\usepackage{bm}
\usepackage{booktabs}
\usepackage{tabularx}
\usepackage{multirow}
\usepackage{epsfig}

\newcommand{\beq} {\begin{equation}}
\newcommand{\eeq} {\end{equation}}
\newcommand{\bea} {\begin{eqnarray}}
\newcommand{\eea} {\end{eqnarray}}
\newcommand{\be} {\begin{equation}}
\newcommand{\ee} {\end{equation}}
\newcommand{\spinup}{\uparrow}

\newcommand{\pll}{\kern 0.56em/\kern -0.8em /\kern 0.56em}

\newcommand{\ud}{\text{d}}

\begin{document}

\title {Spectral properties of 1D extended Hubbard model from bosonization and time-dependent variational principle: applications to 1D cuprates}

\author{Hao-Xin Wang}
\thanks{These authors contributed equally to the work.}
\affiliation{Institute for Advanced Study, Tsinghua University, Beijing, China}
\author{Yi-Ming Wu}
\thanks{These authors contributed equally to the work.}
\affiliation{Institute for Advanced Study, Tsinghua University, Beijing, China}
\affiliation{Stanford Institute for Theoretical Physics, Stanford University, Stanford, California 94305, USA}
\author{Yi-Fan Jiang}
\thanks{jiangyf2@shanghaitech.edu.cn}
\affiliation{School of Physical Science and Technology, ShanghaiTech University, Shanghai 201210, China}
\author{Hong Yao}
\thanks{yaohong@tsinghua.edu.cn}
\affiliation{Institute for Advanced Study, Tsinghua University, Beijing, China}
\date{\today}

\begin{abstract}
Recent ARPES experiments on doped 1D cuprates revealed the importance of effective near-neighbor (NN) attractions in explaining certain features in spectral functions.
Here we investigate spectral properties of the extended Hubbard model with the on-site repulsion $U$ and NN interaction $V$, by employing bosonization analysis and the high-precision time-dependent variational principle (TDVP) calculations of the model on 1D chain with up to 300 sites.
From state-of-the-art TDVP calculations, we find that the spectral weights of the holon-folding and $3k_F$ branches evolve oppositely as a function of $V$.
This peculiar dichotomy may be explained in bosonization analysis from the opposite dependence of exponent that determines the spectral weights on Luttinger parameter $K_{\rho}$.
Moreover, our TDVP calculations of models with fixed $U=8t$ and different $V$ show that $V\approx  -1.7t$ may fit the experimental results best, indicating a moderate effective NN attraction in 1D cuprates that might provide some hints towards understanding superconductivity in 2D cuprates.
\end{abstract}
\maketitle

Since the experimental discovery of high-temperature superconductivity (SC) in cuprates dozens of years ago, there is no consensus yet on the microscopic mechanism for SC in cuprates \cite{Kivelson_Review, Dagotto1994, ZXShen-RMP2003, PALee-RMP2006, Davis2013, Fradkin2015}.
It has been among the most intriguing problems in condensed matter physics and has attracted enormous research efforts, from both experimental and theoretical sides.
Recently, various numerical efforts, including numeric methods such as density matrix renormalization group (DMRG) \cite{White2003, Feiguin2016, Ehlers2017, Dodaro2017, SimonsCollaboration2017, Jiang2018, Hong-Chen2019science, YFJiang-PRR2020, Shoushu2021, Hongchen2021PNAS, Yifan2020, Chung2020, Peng_2021, SimonsCollaboration2020, White2021PNAS, qin2022} and determinant quantum Monte-Carlo (QMC) \cite{White1989, SimonsCollaboration2015,Zixiang2017, Edwin2018}, have been made to address physical properties of two-dimensional Hubbard model and closely related $t$-$J$ models \cite{Anderson1987, Zhang1988, Arovas-ARCMP2022}, which have been believed by many to be the simplest possible models to capture essential physics of 2D cuprates.
However, reliably solving such 2D strongly-correlated problems is extremely challenging from theoretical side and is currently limited by the width of systems in DMRG \cite{Stoudenmire2012} or by the notorious sign problem in QMC \cite{Loh1990, Troyer2005, Wu2005PRB, Li2015PRB, Li2016PRL, Xiang2016PRL, ZXLiQMCreview}.

\begin{figure}
    \centering
    \includegraphics[width=0.48\textwidth]{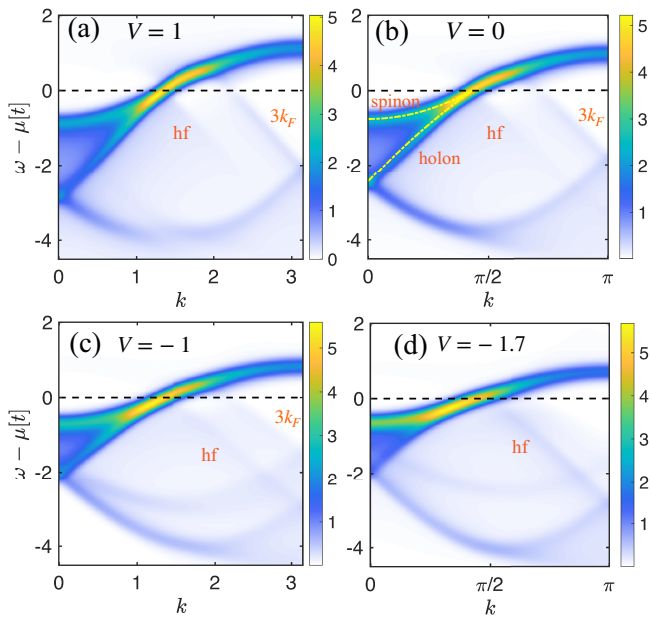}
    \caption{Spectral functions calculated by TDVP for the extended Hubbard model [Eq. \eqref{eq:EHM}] with doping $x=14\%$ and $U=8$ for various values of $V$ between $1$ and $-1.7$ . For $V=0$ in (b), the two yellow dashed lines depict the dispersions for spinon and holon. The holon folding (hf) and $3k_F$ branches are also marked. In (c) and (d), when the NN interaction $V$ is tuned to the attractive side, we find that the $3k_F$ branch becomes more extinguished, in contrast to a sharper hf branch which becomes more visible with sufficiently large $V=-1.7$. }
    \label{fig: color figure}
\end{figure}

One way towards understanding 2D cuprates is to start with 1D cuprates and one important advantage of doing this is the following. 1D microscopic models can be reliably solved such that detailed comparison between theories and experiments can be performed to derive essential interactions for understanding 1D cuprates which are closely related to 2D cuprates.
Indeed, very recently the authors of Ref. \cite{Chen2021} performed an angle-resolved photoemission spectroscopy (ARPES) study of the 1D cuprate material $\text{Ba}_{2-x} \text{Sr}_x \text{Cu O}_{3+\delta}$ with various hole doping and compared their experimental results with predictions of various candidate microscopic models.
Based on the cluster perturbation expansion (CPT) calculations of the extended Hubbard models with 16-site clusters, they found that the experimentally observed spectral features of the holon-folding (hf) branch and the $3k_F$ branch are qualitatively different from those of the 1D standard Hubbard model with only onsite interaction $U$, but can be fitted well when a nearest-neighbor (NN) attractive interaction $V$ is added to the standard Hubbard model (see Fig. \ref{fig: color figure} for details about the hf and $3k_F$ spectral branches).
Attractive interaction between electrons on different sites can be effectively induced by nonlocal Holstein-like electron-phonon couplings (EPC), as shown in recent variational non-Gaussian exact diagonalization (NGSED) studies \cite{wangyao2021, YaoWang2020}.

It remains to be fully understood how an NN attractive interaction $V$ can enhance the hf branch but suppress the $3k_F$ branch simultaneously.
So it is desired to understand the effect of $V$ on electron spectral functions from bosonization, a well-controlled analytical approach in 1D.
It is also desired to numerically study various candidate models on a much longer chain trying to reduce the finite-size effect.
One powerful numerical approach of studying spectral properties of 1D models is the time-dependent variational principle (TDVP), which can efficiently and accurately simulate 1D models on a very long chain.

\begin{figure}
\centering
\includegraphics[width=0.5\textwidth]{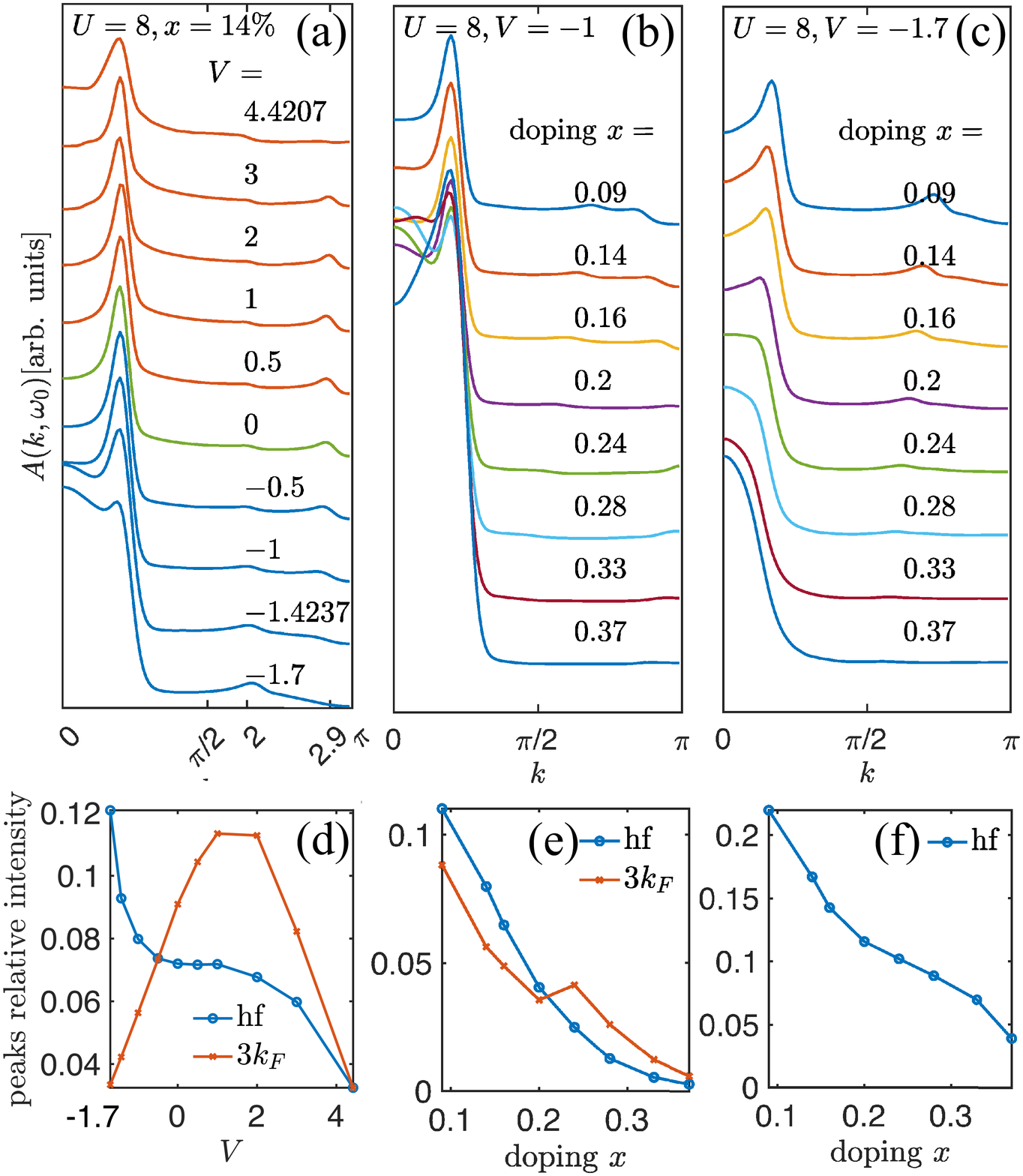}
\caption{Spectral functions of the $U=8$. (a) MDCs under different $V$ with doping concentrate $x=0.14$. The energy cuts $\omega_0$ for difference $V$ range from $-0.7$ to $-2.5$ to match the major peak momenta. (b) MDCs of $V=-1$ at different doping $x$. The energy cuts are taken from $-0.4$ to $-1.1$ for the same reason in (a). (c) MDCs of $V=-1.7$ at different doping $x$.
(d) hf and $3k_F$ peak intensities relative to the major peak as a function of $V$. The peak intensities are extracted from (a). (e-f) hf and $3k_F$ peak relative intensities as a function of  $x$. The peak intensities are extracted from (b) and (c), respectively. }
\label{fig: MDC}
\end{figure}

In this paper, we employ both bosonalization and TDVP to study the 1D extended Hubbard model [Eq. (\ref{eq:EHM})], with emphasis on understanding how peak features in the electron spectral function varies with NN interaction $V$ as well as doping $x$.
From large-scale TDVP calculations of the extended Hubbard model on a 1D chain with up to 300 sites, we obtained the spectral function for various values of $V$ and $x$, while fixing $U=8t$ and $t=1$.
Our TDVP simulations showed that an attractive $V$ can enhance the hf branch but suppress the $3k_F$ branch while a repulsive $V$ would reverse the trend.
Our numerical results, as shown in Fig. \ref{fig: color figure} and Fig. \ref{fig: MDC}, are qualitatively consistent with ones from previous studies \cite{Chen2021,wangyao2021, YaoWang2020}.
In the following, we shall present the results of our large-scale TDVP calculations and then provide analytical explanations from bosonization for such dichotomy between repulsive and attractive $V$.

{\it Model}.---The Hamiltonian of 1D extended Hubbard model is given as
\begin{equation}
  \begin{aligned}
    H=&-t\sum_{\braket{ij},\sigma}c_{i\sigma}^{\dagger}c_{j\sigma}
    +U\sum_i n_{i\spinup}n_{i\downarrow}+V\sum_{\langle i j \rangle} n_i n_j,\label{eq:EHM}
  \end{aligned}
\end{equation}
where $\braket{ij}$ denotes NN sites, $\sigma = \uparrow$/$\downarrow$, $n_{i\sigma} = c_{i\sigma}^{\dagger} c_{i\sigma}$ and $n_i = n_{i\uparrow} + n_{i\downarrow}$. In 1D cuprate, $t\simeq 0.6 \mathrm{eV}$ and $U\simeq 8t$ \cite{hybertsen1990,Chen2021}. Hereafter for simplicity we fix $U=8t$ and set $t=1$ as unit of energy. For electrons with Coulomb interactions, one naturally expects that the NN interaction $V$ is repulsive.
But, electron-phonon coupling (EPC) can generate an effective attractive NN interaction with retardation.
Thus, we consider both cases of positive and negative $V$ in our calculations.
Thanks to the particle-hole symmetry of the extended Hubbard model, we only need to focus on the case of hole doping.
The number of sites, electrons, and doping concentration are denoted by $L$, $N$, and $x = 1-N/L$, respectively.

{\it Spectral function from TDVP.}---Over the past few years, much progress has been made in developing controllable numeric methods based on tensor-network states \cite{Orus2014, Cirac2021Review, Evenbly2022, Orus2019, Verstraete2008}.
Among others, the time-dependent variational principle (TDVP) is currently state-of-the-art approach to calculate dynamic correlations in 1D, thus being widely used to compute the spectral function in strongly correlated 1D systems \cite{Verstraete2011TDVP, Verstraete2013TDVP, Verstraete2016TDVP, tian2021, Li2022, Xu2022, Kloss2018, Yantao2020, Mingru2020, Secular2020, Hemery2019, Paeckel2019}.
The basic idea of TDVP is to project the time evolution of the wave function to the tangent space of the matrix product state (MPS) sub-manifold, and offer the optimal way of the truncation step.
According to the entanglement structure in 1D correlated systems, the MPS methods are generally more reliable than perturbation methods; the leading error of the former comes from the truncation error, which can be overcome by increasing  bond dimensions.
Here we implement the finite-size TDVP algorithm to calculate the retarded Green's function of electrons $G^R(x,t)$ and then obtain the spectral function $A(k,\omega)$ by performing a double Fourier transform.
We analyzed the finite-size effect to ensure the robustness of our numerics, and the details can be found in the SM \cite{Supplementary}.

In Fig. \ref{fig: MDC} we show the calculated $A(k,\omega)$ from TDVP for the extended Hubbard model with various $V$ and doping concentration $x$ on system size $L=100$. Fig.~\ref{fig: MDC}(a) shows the momentum distribution curves (MDCs), i.e. $A(k,\omega)$ at some specific $\omega=\omega_0$, of the extended Hubbard model with different $V$ (we plot $V>0$ and $V<0$ cases in different colors). For the pure Hubbard model with $V=0$, the energy cut is taken as $\omega_0 = -1.3$ measured from Fermi energy. For the other values of $V$, we gradually shift $\omega_0$ such that the hf and the $3k_F$ branches are most visible at momenta $k\simeq 2$ and $k\simeq 2.9$ respectively. The smallest value of $V = -1.7$ is close to the critical boundary and for $V\lesssim -1.8$ phase separation start to appear in the ground state of the EHM.

It is clear from Fig.~\ref{fig: MDC}(a) that as $V$ decreases from $1$ to $-1$, the hf spectral weight stays almost intact; as $V$ decreases further, the increases of hf become evident.
Meanwhile, the $3k_F$ holon weight decreases monotonically. For $V>1$, as $V$ increases, the hf weight changes insignificantly, while the $3k_F$ weights suddenly drop at $V\approx V^*\equiv -U/(2\cos(2k_F))$ (see SM \cite{Supplementary} for detailed discussion).
To clearly characterize the variation of these two weights, we extract the relative peak intensity, which is defined as the hf or $3k_F$ peak intensities divided by the major intensity, in Fig.~\ref{fig: MDC}(d).
Here we do not use the Lorentzian peak fitting but directly use the spectral weight.
Both proper repulsive and attractive $V$ can largely suppress $3k_F$ weight, but hf weight decays monotonically as $V$ increases.

The dichotomous feature of $k_F$ holon-folding and $3k_F$ holon spectral weight in the presence of $V$ can be used to diagnose the interaction nature in 1D cuprates. In accord with the spectral behaviors reported in Ref. \cite{Chen2021}, the observed spectral weight of hf is fully evident but that of the $3k_F$ branch disappears, which leads the authors to conclude that some NN attraction must exist in 1D BSCO material. On the aspects of our results, although both proper repulsive and attractive $V$ can largely suppress $3k_F$ weight, repulsive $V\simeq 4.4$ is usually regarded as unrealistic in doped cuprates since the Coulomb repulsion is screened and becomes quite local.  Our results confirm only a large enough attractive $V$, e.g., $V=-1.7$, can be consistent with experiment features.

We show the doping dependent MDCs with $U=8$ at two different values of $V=-1$ and $-1.7$ in Fig. \ref{fig: MDC}(b,c), and the corresponding peak relative intensities are shown in Fig. \ref{fig: MDC}(e,f).
The cases of $U=8, V=-1$ at different doping have been discussed in Ref. \cite{Chen2021}. For $V=-1$ we find that the intensities of hf and $3k_F$ are comparable. However, for $V=-1.7$, the $3k_F$ intensity gets totally smeared out, and the hf weight decays more rapidly over increasing doping. As the two features of $V=-1.7$ are consistent with the recent experimental results, we think that the 1D cuprates might have a relatively strong NN attraction around $V=-1.7$.

{\it Phenomenological bosonization.}---Bosonization is a powerful and reliable method in analyzing low-energy properties of 1D correlated models.
In low-energy limit, the linearized 1D dispersion allows for an exact identity relating the original fermion operators with bosonized field operators \cite{Mandelstam, Heidenreich1980, Luther, Mattis}, and the interacting system of fermions may be turned into a free theory of bosons.
However, in the higher energy scale, the nonlinearity of the dispersion turn on interactions between bosons \cite{Haldane_1981}, which impedes our way of obtaining fermion properties by bringing about additional complexities in evaluating the expectation of boson field exponentials in an interacting bosonic theory.

Here we follow the seminal work by Haldane and adopt his phenomenological bosonization \cite{Haldane1981prl}. We start with the density operator, $\rho_s(x) = \sum_i\delta(x-x_i)  = \sum_n |\nabla \phi_s(x)| \delta(\phi_s(x)- 2\pi n)$, where $\phi_s(x)$ is a monotonically increasing function of position, which takes the value $\phi_s(x_i)= 2\pi i $ at the position of the $i$th electron with spin polarization $s=\pm$. Here $\phi_s(x)$ is connected with the boson field
$\Phi_s(x)$ by $\phi_s(x) = \pi \rho_0 x-\Phi_s(x)$ where $\Phi_s(x)=[\Phi_\rho(x)+s\Phi_\sigma(x)]/\sqrt{2}$ with the subscript $\rho$ and $\sigma$ labeling the charge and spin sector.
Using the Poisson summation formula, the density operator can be rewritten as
\begin{eqnarray}
    \rho_s(x) = \left[ \rho_0 - \frac 1 \pi \nabla \Phi_s(x)\right] \sum_m \exp\left[2im(\pi \rho_0 x -\Phi_s(x) \right].~~~~
\end{eqnarray}
The fermion fields are followed by taking the square root of $\rho(x)$, and introducing another field $\Theta(x,t)$ to ensure the fermion anti-commutation rule. Explicitly we have
\begin{equation} \label{eq: bosonization}
  \begin{aligned}
    &\Psi_{s}(x,t)\sim\sum_{m=-\infty}^{\infty}\Big\{\exp(ic_m k_F x)\\     & \exp[i\Theta_\rho(x,t)/\sqrt{2}]\times
 \exp[ic_m\Phi_\rho(x,t)/\sqrt{2}] \\
 & \exp[is\Theta_\sigma(x,t)/\sqrt{2}]\times
 \exp[isc_m\Phi_\sigma(x,t)/\sqrt{2}]\Big\}
  \end{aligned}
 \end{equation}
where $k_F$ is the Fermi momentum, $c_m=2m+1$ with $m$ an integer. The leading harmonics is given by $c_m=\pm1$, representing the right and left movers respectively. If only these two  are kept, we arrive at the exact bosonization formula, for which the Hamiltonian is given by $H_0=\frac{1}{L}\sum_{\nu,k}\left(\frac{\pi v_\nu K_\nu}{2}\Pi_\nu(k)\Pi_\nu(-k)+\frac{v_\nu}{2\pi K_\nu}k^2\Phi_\nu(k)\Phi_\nu(-k)\right)$ with $\Pi_\nu(x)=\nabla\Phi_\nu(x)/\pi$ \cite{JVoit_1995}.
Upon including higher harmonics, interacting terms between boson fields are effectively generated. However, we can assume that these interactions are weak, i.e. the nonlinearity of the original fermion dispersion is small as long as we are still in a low energy scale.
Then we can evaluate the single-particle Green's function using $H_0$. For instance, the retarded Green's function for the spin-up fermions is given by
$G^{R}_{\spinup}\left(x, t; x^{\prime}, t^{\prime}\right) \equiv -i \theta\left(t-t^{\prime}\right)\left\langle\left\{\Psi_{\spinup}(x, t), \Psi_{\spinup}^{\dagger}\left(x^{\prime}, t^{\prime}\right)\right\}\right\rangle_{H_0}$.
A straightforward derivation shows that (see \cite{Supplementary} for details) for spacetime translation invariant system
 $G^R_{\spinup}(x,t;0,0)=\sum_m G^R_{\uparrow, (2m+1)k_F}(x,t)$, where
 \begin{eqnarray}
     &&G^R_{\uparrow, (2m+1)k_F}(x,t)\sim-\theta(t)e^{i c_m k_F x}\times\nonumber\\
     &&~~\text{Re} \prod_{\nu=\rho,\sigma}\frac{1}{[\alpha+i(u_\nu t-x)]^{c_m/2}}\left[\frac{\alpha^2}{(\alpha+iu_\nu t)^2+x^2}\right]^{\gamma_{\nu,m}}, ~~~
     \label{eq:GR}
 \end{eqnarray}
 where $\alpha$ is a cutoff with the scale of lattice constant. The power index $\gamma_{\nu,m}$ is given by
 \begin{equation}
   \gamma_{\nu,m}=\frac{1}{8}\left(c_m^2 K_\nu+\frac{1}{K_\nu}-2c_m\right),\label{eq:gamma}
 \end{equation}
where $K_{\nu=\sigma,\rho}$ are the Luttinger parameters that can be extracted from numerical calculations (see below).
For $m=0$, the Green's function reduces to the well-documented results of $k_F$ branch \cite{JS, Voit1998, Suzumura, Schulz_1983}, including the structure of spinon, holon, holon-folding, and anti-spinon, with different velocity $\pm u_{\rho/\sigma}$. The Green's function of $3k_F$ branch also reproduces the earlier analysis \cite{Ren_Anderson} with  $K_\rho=1/2$.

To obtain the spectral function $A(k,\omega)$, one needs to perform a 2d Fourier transformation on the retarded Green function \eqref{eq:GR}, which is, unfortunately, a rather involved task given the complexity of the function structure. Here, we focus on the singular behavior of $A(k,\omega)$ since they dominantly characterize the excitations. The spectral function of the $k_F$ branch ($m=0$) has been studied in previous work \cite{orignac2011, meden1992, Voit_1993}. A straightforward generalization to $m\ne 0$ leads to the singular spectral weight  near the excitation dispersion,
\begin{equation} \label{A singularity}
\begin{aligned}
    A_m(c_m k_F+q, \omega) \sim& | \omega - u_\rho q |^{2\gamma_{\sigma,m}+\gamma_{\rho,m} - 1/2} \\
  &\times  | \omega - u_\sigma q |^{2\gamma_{\rho, m}+\gamma_{\sigma,m} - 1/2}\\
  &\times  |\omega + u_\rho q|^{\gamma_{\rho,m} + 2\gamma_{\sigma,m}} .
    \end{aligned}
\end{equation}
The spectral function in the above expression diverges at $\omega = \pm u_{\nu} q$. Nevertheless, divergences do not present in both experiments and numeric data.
To take a regular value and investigate the model dependence of the spectral weights, we cut off the spectral at $|\omega \pm u_{\nu} q| > \delta$ with $\delta$ a small positive value. It then brings about the peak intensities are a monotonously decreased function of indices $\gamma_{\rho,m}$.

\begin{table}[tbp]
    \centering
    \begin{tabular}{lrrrrrr}
    \hline  \hline
       $V$  & $0.5$ & $0$ & $-0.5$ & $-1$ & $-1.4237$ & $-1.7$  \\
       \hline
        $K_\rho$  &  0.541 & 0.577& 0.620&  0.694& 0.814& 0.933 \\
       $\gamma_{\rho,0}$ (hf) & 0.0487 &  0.0389   &   0.0292    &  0.0169  &   0.0053    &0.0006   \\
       $\gamma_{\rho,1}$ ($3k_F$) & 0.0897   &   0.1155  &   0.1488  &    0.2108   &  0.3195    &      0.4335\\
\hline  \hline
\end{tabular}
    \caption{The Luttinger parameter $K_\rho$ together with the corresponding indices $\gamma_{\rho, m}$ of EHM as the function of NN interaction $V$.  $\gamma_{\rho,0}$ and $\gamma_{\rho,1}$ control the intensities of hf and $3k_F$, respectively. The higher value of $\gamma$'s the greater the intensities, and vice versa. }
    \label{tab: Kc and gamma}
\end{table}

Now we apply Eq.\eqref{A singularity} to the microscopic models. Since the model preserves the spin $SU(2)$ symmetry, we have $K_\sigma=1$ and hence $\gamma_{\sigma,m}$ remains constant. The only chance that $V$ affects the spectral function is through $K_\rho$, or $\gamma_{\rho,m}$ equivalently. We then focus on the contributions related to $\gamma_{\rho,m}$. The hf and $3k_F$ branches are given by
\begin{equation}
\begin{aligned}
   \text{holon-folding:} \quad & A_0(k_F+q, \omega) \sim |\omega + u_\rho q|^{\gamma_{\rho,0}},\\
   3k_F: \quad & A_1(3k_F+q, \omega) \sim |\omega - u_\rho q|^{\gamma_{\rho,1}}.
\end{aligned}
\end{equation}
Here, we deploy the density matrix renormalization group (DMRG) calculation to extract $K_\rho$ via the charge structure factor accurately. In a charge gapless phase, the charge structure factor has the form of $S_c(k) = K_\rho k/\pi $ \cite{Daiwei2022}. We show the values of $K_\rho$ as a function of $V$ and the corresponding indices $\gamma_{\rho,m}$ with $m = 0,1$, as defined in Eq.\eqref{eq:gamma}, in Table~\ref{tab: Kc and gamma}. We find as $V$ decreases from $0.5$ to $-1.7$, the Luttinger parameter $K_{\rho}$ increases to nearly $1.0$. At the same time,  $\gamma_{\rho,0}$ decreases while $\gamma_{\rho,1}$ increases significantly. As a result,
the hf intensity will be enhanced while the $3k_F$ intensity will be suppressed largely as $V$ decreases, consistent with our previous numeric observations.

{\it Summary and discussion.}---In this study, we investigated the spectral properties of the 1D extended Hubbard model, partly inspired by the recent ARPES experiment on the 1D cuprate BSCO \cite{Chen2021}. In particular, we provide both analytical and numerical evidences of how the holon-folding and $3k_F$ spectral weights vary with the NN density interaction $V$ and doping $x$. When compared to the experimental data, our results suggest that an attractive $V\approx -1.7$ can fit well with the experimental results in 1D BSCO.

It is interesting to note that our numerical results suggest that the NN attraction is approximately $V\approx -1.7$, which is slightly larger than that from previous predictions ($V\approx -1$) in \cite{Chen2021} to best fit with the experimental observations.
This quantitative difference may result in a change the ground state at, e.g. quarter filling, from the Luttinger liquid, when $V\approx -1$, to the Luther-Emery liquid with a dominant pair-pair correlation, when $V\approx -1.7$ \cite{Daiwei2022}. Furthermore, it has been found that superconductivity from the extended Hubbard model is enhanced with the increase of the NN attraction $V$ \cite{Peng2022}. With the help of a moderate NN attraction in cuprates as our calculations suggested in this work, superconductivity may become dominant over the charge density wave when generalizing the 1D cuprates to 2D ones.

{\it Acknowledgement.} This work is supported in part by the NSFC under Grant No. 11825404 (H.-X.W., Y.-M.W. and H.Y.), the MOSTC under Grant Nos. 2018YFA0305604 and 2021YFA1400100 (H.Y.), the CAS Strategic Priority Research Program under Grant No. XDB28000000 (H.Y.), Shanghai Pujiang Program under Grant No.21PJ1410300 (Y.-F.J.). Y.-M.W. was supported by Shuimu Fellow Foundation at Tsinghua and also acknowledges the Gordon and Betty Moore Foundation’s EPiQS Initiative through GBMF8686 for support at Stanford.

{\it Note added.}---While finishing this work, we noticed a related and interesting work \cite{Tang2022} that studied the dynamic properties of the Hubbard-extended-Holstein model and the extended-Hubbard model. Consistent results  are obtained when overlapping occurs in these two works.

%\nocite{*}

\widetext

%\pagebreak
%\begin{center}
\section{Supplemental Materials}
%\end{center}

\setcounter{equation}{0}
\setcounter{figure}{0}
\setcounter{table}{0}
%\makeatletter
\renewcommand{\theequation}{S\arabic{equation}}
\renewcommand{\thefigure}{S\arabic{figure}}
\subsection{A. Numeric calculation of single-particle spectral function}

We implement the finite-size TDVP to calculate the single-particle retarded Green's function, which is defined as
\begin{equation}
	G_{\sigma \bar\sigma}^{R}\left(x, t ; x^{\prime}, t^{\prime}\right) \equiv-i \theta\left(t-t^{\prime}\right)\left\langle\left\{c_\sigma(x, t), c_{\bar\sigma}^{\dagger}\left(x^{\prime}, t^{\prime}\right)\right\}\right\rangle,
\end{equation}
where $\{A, B\} = AB+BA $ is the anti-commutator. After that, a double Fourier transform is applied to find the Green's function in momentum-frequency space $G^R_{\sigma, \bar \sigma}(k,\omega)$. The single-particle spectral function is followed by the imaginary part of the Green's function, namely,
\begin{equation}
    A(k,\omega) = -\frac{1}{\pi} \sum_{\sigma, \bar\sigma} \text{Im}G^R_{\sigma, \bar \sigma}(k,\omega).
\end{equation}

In practice, the TDVP code is imposed with the total particle number and total spin $S_z$ conservation. When calculating the retarded Green's function, the Fermion operator at time slice $t=0$ is fixed on the middle of the chain, by assuming the translation invariance\cite{tian2021}. The bond dimensions of MPS are retained $D=500\sim 2000$,  which gives the truncation error magnitudes about $10^{-7}$.
We take open boundary conditions in accordance with the MPS' preference. The step size $\tau = 0.02 \sim 0.03$, and the maximum time $t_{\text{max}} = 40 \sim 90$. The step error of such small $\tau$ can be neglected, and such a long time of correlation supports us to obtain the spectral functions brute-force, without any kind of extrapolation or prediction.
In addition, when performing the Fourier transform $t\to\omega$, we multiply a Gaussian window $e^{-\alpha t^2}$ on the integral, with $\alpha$'s magnitude about $0.01$. The aforestated parameters support a reliable dynamic simulation with high resolution.

As mentioned above, we assume the translation invariance and fix the Fermion operator at $t=0$ in the middle of the chain, which could help us largely reduce the simulation cost.  However, the assumption only comes to be true in the thermodynamic limit $L\to \infty$. In order to plug up the loophole, we calculate the spectral function on different sizes and see the finite-size scaling. Figure~\ref{fig: color figure}(b) shows the momentum distribution curves (MDCs) at $\omega - \mu = -1$ of the Hubbard model, with $U=6$ and $x=0.14$. The MDCs of different sizes collapse indistinguishably with the naked eye. Thus, We take the system size $L = 100$ in most of the following calculations and regard the data as in thermodynamic limit.
\begin{figure}
\centering
\includegraphics[width=0.6\textwidth]{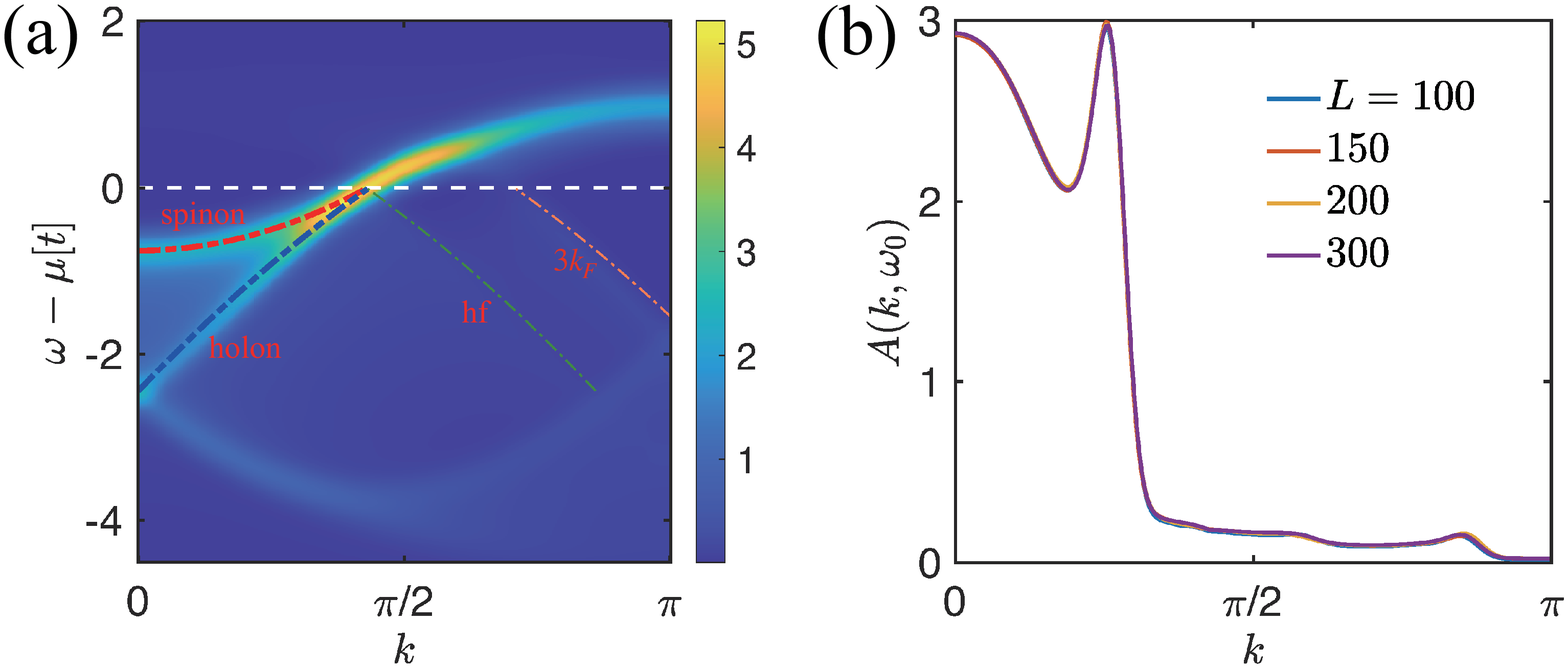}	
\caption{(a) The color-scale plot of the spectral function of the Hubbard model at $U=8$, length $L=100$ and doping level $x=0.14$. Only half of the Brillouin zone is drawn considering the parity symmetry. The white dashed line denotes the Fermi energy as a guide. The dash-dotted lines are the guides for the dispersion of low-energy excitations, including the spinon, holon, hf, and $3k_F$ branches.  (b) The MDCs finite-size scaling of the Hubbard with $U=6$ and doping $x=0.14$, at energy cut $\omega - \mu = -1$. The curves of different sizes collapse well so that $L=100$ is sufficiently large to represent the thermodynamic limit behavior.
}
\label{SMfig: color figure}
\end{figure}

\subsection{B. Phenomenological bosonization calculation of the Green's function}
There are various ways to obtain the single particle correlation functions. A very convenient way is utilizing the boson coherent path integral. However, such an approach has to be supplemented by analytic continuation, which in fact does not ease the calculation. Here we evaluate the correlation function directly in real-time space, and below as the explicit steps to obtain \eqref{eq:GR}.
The first step is to diagonalize the Hamiltonian $H_0=\sum_\nu H_\nu$ using Bogoliubov transformation. After Fourier transformation into momentum space, we have
\begin{equation}
  H_\nu=\frac{1}{L}\sum_k\left(\frac{\pi v_\nu K_\nu}{2}\Pi_\nu(k)\Pi_\nu(-k)+\frac{v_\nu}{2\pi K_\nu}k^2\Phi_\nu(k)\Phi_\nu(-k)\right)\label{eq:H2}
\end{equation}
Introducing
\begin{equation}
  \begin{aligned}
    \gamma_k=\frac{1}{\sqrt{2}}\left(\sqrt{\frac{|k|}{\pi K_\nu}}\Phi_\nu(k)+i\sqrt{\frac{\pi K_\nu}{|k|}}\Pi_\nu(k)\right),~ \gamma_k^\dagger=\frac{1}{\sqrt{2}}\left(\sqrt{\frac{|k|}{\pi K_\nu}}\Phi_\nu(-k)-i\sqrt{\frac{\pi K_\nu}{|k|}}\Pi_\nu(-k)\right),
  \end{aligned}
\end{equation}
it is easy to verify that these two operators obey $[\gamma_k,\gamma_{k'}^\dagger]=\delta_{k,k'}$. The inverse transformation is also easy to obtain,
\begin{equation}
  \Phi_\nu(k)=\sqrt{\frac{\pi K_\nu}{2|k|}}(\gamma_k+\gamma_{-k}^\dagger), ~ \Pi_\nu(k)=-i\sqrt{\frac{|k|}{2\pi K_\nu}}(\gamma_k-\gamma_{-k}^\dagger)\label{eq:inverse_bogoliubov}
\end{equation}
Substitute \eqref{eq:inverse_bogoliubov} to \eqref{eq:H2}, the diagonalized Hamiltonian is obtained
\begin{equation}
  H_\nu=v_\nu\sum_k|k|\gamma_k^\dagger\gamma_k\label{eq:H3}
\end{equation}
Under this Hamiltonian, the time evolution of the free operator $\gamma_k$ is simple, namely $\gamma_k(t)=\gamma_k e^{-i v_\nu|k|t}$.

Next, we evaluate the Green's function using the diagonalized Hamiltonian \eqref{eq:H3}. Notice that the charge and spin sectors are decoupled, this fact enables us to write $G^>(x,t)=\braket{\Psi^\dagger_{+,\spinup}(x,t)\Psi_{+,\spinup}(0,0)}$ as a product of charge and spin contributions:
\begin{equation}
  G^>(x,t)=\prod_\nu\left\langle\exp\left(\frac{-i}{\sqrt{2}}(c_m\Phi_\nu(x,t)-\Theta_\nu(x,t))\right)\exp\left(\frac{i}{\sqrt{2}}(c_m\Phi_\nu(0,0)-\Theta_\nu(0,0))\right)\right\rangle_{H_\nu}
\end{equation}
where the subscript $H_\nu$ means average under $H_\nu$. For free boson Hamiltonians like \eqref{eq:H2}, the boson field is bilinear (Gaussian), and one can make use of the fact that the second order cumulant expansion is exact, thus we evaluate the average using $\langle e^{iA}e^{-iB}\rangle=\exp(-\frac{1}{2}\braket{A^2}-\frac{1}{2}\braket{B^2}+\braket{AB})$. For the charge contribution, we have
\begin{equation}
G^>_c(x,t)=\exp\left[\frac{1}{2}\braket{(c_m\Phi_c(x,t)-\Theta_c(x,t))(c_m\Phi_c(0,0)-\Theta_c(0,0))}-\frac{1}{2}\braket{(c_m\Phi_c(0,0)-\Theta_c(0,0))^2}\right]\label{eq:G_greater}
\end{equation}
In the exponential there are two parts, but we can skip the calculation of the second term since it can be obtained from the first term by taking the limit of $(x,t)\to(0,0)$. The operators in the first term, when written in momentum space is
\begin{equation}
  \begin{aligned}
    &\langle(c_m\Phi_c(x,t)-\Theta_c(x,t))(c_m\Phi_c(0,0)-\Theta_c(0,0))\rangle\\
    =&\frac{1}{L^2}\sum_{k,k'} e^{ikx}\left\langle\left(c_m\Phi_c(k,t)-\frac{\pi\Pi_c(k,t)}{ik}\right)\left(c_m\Phi_c(-k')+\frac{\pi\Pi_c(-k')}{ik'}\right)\right\rangle\\
    =&\frac{\pi}{L^2}\sum_{k,k'} e^{ikx} \left\langle\left[c_m\sqrt{\frac{K_c}{2|k|}}(\gamma_k+\gamma^\dagger_{-k})+\sqrt{\frac{|k|}{2K_c}}\frac{1}{k}(\gamma_k-\gamma^\dagger_{-k})\right]\left[c_m\sqrt{\frac{K_c}{2|k'|}}(\gamma_{-k'}+\gamma^\dagger_{k'})-\sqrt{\frac{|k'|}{2K_c}}\frac{1}{k'}(\gamma_{-k'}-\gamma^\dagger_{k'})\right]\right\rangle\\
    =&\frac{1}{2}\int_0^\infty dk\frac{e^{ikx}}{2k}\left(c_m^2K_c+\frac{1}{K_c}+2c_m\right)e^{-iv_c|k|t}+\frac{1}{2}\int_0^\infty dk\frac{e^{-ikx}}{2k}\left(c_m^2K_c+\frac{1}{K_c}-2c_m\right)e^{-iv_c|k|t}\label{eq:part1}
  \end{aligned}
\end{equation}
To obtained the last line, we have used the fact that at $T=0$, the Bose distribution function $\braket{\gamma_k^\dagger\gamma_k}$ vanishes. The remaining integrals over $k$ run from $0$ to $\infty$. As we put in the main text that there exists an energy cut off beyond which the bosonization procedure fails, thus for consistency the upper limit of the momentum integral cannot be taken formally as $\infty$. The cutoff is made by hand, and this can be done by introducing an exponential decaying factor $e^{-\alpha k}$ into the integrand and the small parameter $\alpha$ has the same order of lattice constant. Collecting the results in \eqref{eq:part1}, the exponential in \eqref{eq:G_greater} becomes,
\begin{equation}
  \begin{aligned}
     &-\frac{1}{4}\left[\left(c_m^2K_\nu+\frac{1}{K_\nu}\right)\int_0^\infty dk \cdot e^{-\alpha k} \frac{1-\cos(kx)e^{-i u_\nu|k|t}}{k}-i 2c_m\int_0^\infty dk \cdot e^{-\alpha k}\frac{\sin(kx)}{k}e^{-i u_\nu|k|t}\right]\\
     &=\frac{1}{8}\left[\left(c_m^2K_\nu+\frac{1}{K_\nu}\right)\ln\frac{\alpha^2}{(\alpha+i u_\nu t)^2+x^2}-2c_m\ln\frac{\alpha+i(u_\nu t-x)}{\alpha+i(u_\nu t+x)}\right]
  \end{aligned}
\end{equation}
which, in combination with $G^<(x,t)$, immediately gives \eqref{eq:GR}.

\subsection{C. Microscopic origin of $3k_F$ branch.}
\begin{figure}
    \centering
    \includegraphics[width=0.5\textwidth]{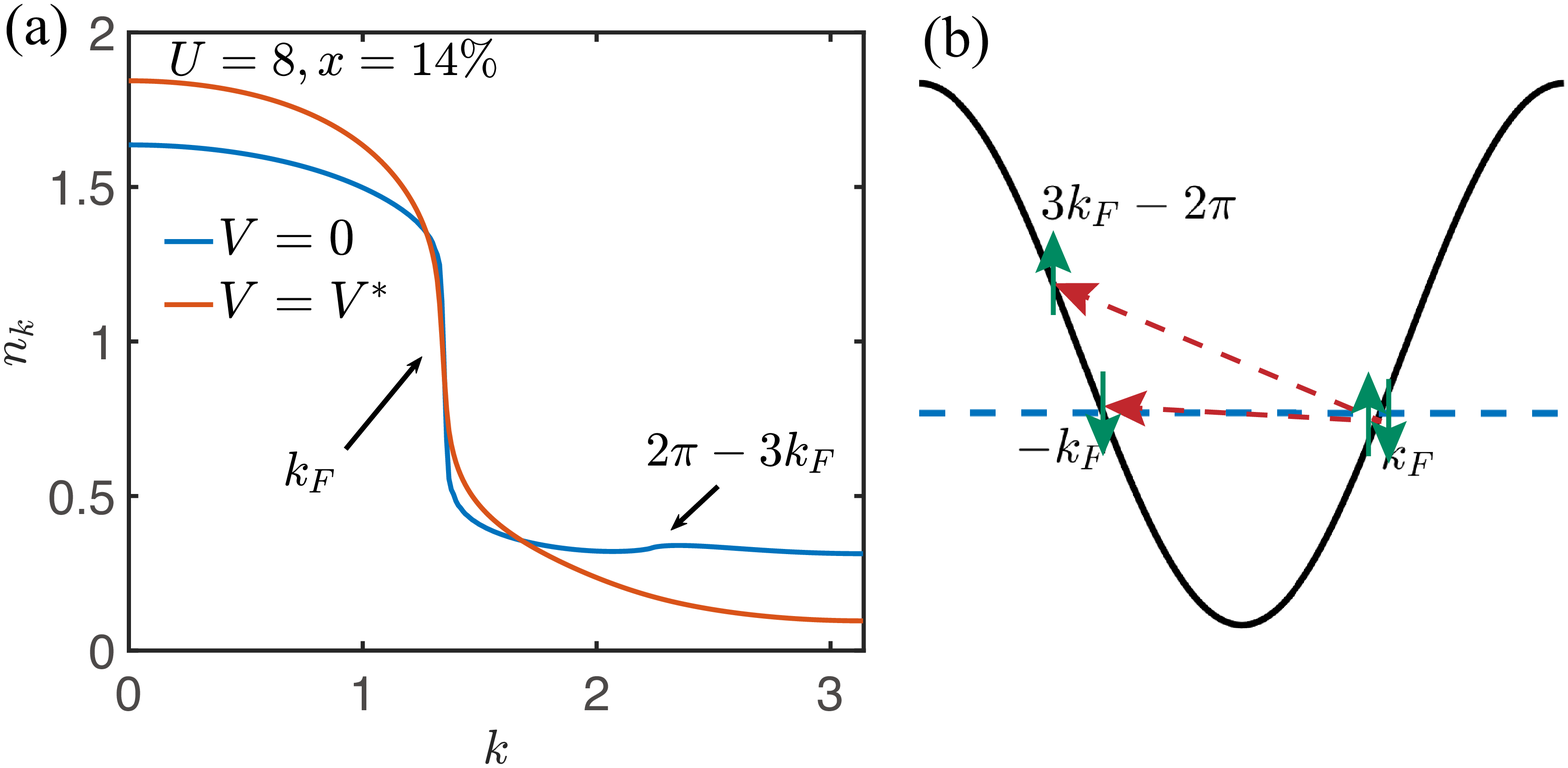}
    \caption{(a) The charge density distribution $n_k$ of EHM under two different $V=0$ and $V=V^*\simeq 4.4207$. We choose $U=8$ and doping level $x=14\%$.
    	The black arrows point to singularities associated with excitation modes emanating from different momenta. A hump around $k=2\pi-3k_F$ appearing at $V=0$ corresponds to the $3k_F$ excitation.
    	Under this special value of $V = V^*$, the singularity at $k = 2\pi - 3k_F$ disappears, implying the elimination of the $3k_F$ excitation.
    	(b). The illustration of the umklapp $g_{3,\perp}$ process.  The black curve is the dispersion and the blue dashed line is the Fermi energy. two right-moving particles (the green arrows) with opposite spin polarization around the Fermi surface are scattered to the left-moving side,  one of which acquires momentum $k = 3k_F - 2\pi$.}
    \label{fig: nk}
\end{figure}

We see from above that the $3k_F$ spectral weight can also be suppressed with a large repulsive $V$. Below we show that this peculiar behavior is closely related to the origin of the $3k_F$ branch.
To address this issue, we note that the low-energy excitation modes are closely related to a singular behavior in the charge density distribution $n_k = \sum_{s=\uparrow,\downarrow} c_{k,s}^\dagger c_{k,s}$.
For example, the holon/spinon excitation around $k_F$ gives rise a dip of $n_k$ at the Fermi momentum $k_F$. Similarly, the $3k_F$ mode contributes to a hump structure in $n_k$ at $k = 2\pi - 3k_F$.  In Fig.~\ref{fig: nk}(a), the blue curve shows these behaviors of $n_k$ obtained by performing DMRG calculations for the Hubbard model with $U=8$ and $V=0$ at doping $x=14\%$. Note that the singularity behaviors of $n_k$ are directly connected to the spectral function via an integral.

It may be contrary to the common sense why the particle number does not monotonically decrease as the band energy increases in the sense that the particles usually prefer lower energy states. The hump in high energy implies it may come from a gapped process, i.e. the umklapp scattering in the Hubbard model away from half-filling, which can be written as
\begin{eqnarray}
H_3 = \int \ud x \sum_{\sigma =\uparrow,\downarrow,s = L, R}
g_{3,\perp} \Psi^\dagger_{s,\sigma}  \Psi^\dagger_{s,-\sigma} \Psi_{-s,-\sigma} \Psi_{-s,\sigma},
\end{eqnarray}
where the argument $x$ of the fermion field operators $\Psi_{s,\sigma}(x)$ is omitted for brevity.
The process $g_{3,\perp}$ is schematically shown in Fig.~\ref{fig: nk}(b).  Obviously, the scattering transfers the particles around the Fermi surfaces to high energy states with momentum $3k_F$, resulting in the singular hump in the particle number distribution $n_k$ at $3k_F$.
We numerically verified the above interpretation of origination of $3k_F$ in the EHM model, where $g_{3,\perp} = U+2V\cos(2k_F)$.
By increasing $V$ to a critical value $V^*$ which satisfies $U+2V^*\cos(2k_F) = 0$, we expect that $g_{3,\perp}$ vanishes and hence the umklapp  process is canceled, and as a result, the $3k_F$ mode disappears. We show $n_k$ of the EHM model with $U = 8, V = V^*$ and $x = 14\%$ in Fig.~\ref{fig: nk}(a). Clearly, the peak around $3k_F$ is smeared out, indicating the absence of $3k_F$.

\end{document}